\documentclass[11pt,twoside]{article}
\usepackage{asp2006}
\usepackage{epsf}
\usepackage{psfig}
\usepackage{graphics}
\usepackage{lscape}
\usepackage{natbib}
\usepackage{amssymb}

\begin{document}

\title{Inflow of the Broad-Line Region and the Fundamental Limitations of Reverberation Mapping}

\author{C. Martin Gaskell}
\affil{Astronomy Department, University of Texas, Austin, TX 78712-0259, USA}

\begin{abstract}
The evidence from velocity-resolved reverberation mapping showing a net infall of the broad-line region (BLR) of AGNs is reviewed.  Different lines in many objects at different epochs give a consistent picture of BLR motions.  The motions are dominated by virialized motion (rotation plus turbulence) with significant net inflow.  The BLR mass influx is sufficient to power the AGN.  The increasing blueshifting of lines with increasing ionization potential is a consequence of scattering off infalling material.  The high blueshiftings of the UV lines in Narrow-Line Seyfert 1s are due to enhanced BLR inflow rates rather than strong winds.  Seemingly conflicting cases of apparent outflow reverberation mapping signatures are a result of the breakdown of the axial-symmetry assumption in reverberation mapping.  There are several plausible causes of this breakdown:  high-energy variability tends to be intrinsically anisotropic, regions of variability are necessarily located off-axis, and  X-ray observations reveal major changes in line-of-sight column densities close to the black hole.  Results from reverberation mapping campaigns dominated by a single event need to be treated with caution.

\end{abstract}

\section{Introduction}

At a conference on ``Accretion and Ejection in AGNs'' one could na\"ively expect roughly equal time for both accretion and ejection!  However, inspection the AGN literature (and the conference program) shows that, while the theory of ejection in jets and winds, and observations of ejection from $\gamma$-rays to radio waves are all well covered, there is almost nothing on observations of {\em inflow}.  Yet we all believe that AGNs are ultimately powered by accretion onto black holes, so {\em something} has to be inflowing!  I argue here that the ``something'' is the broad-line region (BLR) and that we can see consequences of the inflow.

\section{The broad-line region -- a ``bird's nest''}

A BLR is a ubiquitous feature of all AGNs accreting at a high accretion rate (within two or three orders of magnitude of the Eddington limit).  The lines are the dominant features of the UV/optical spectrum.  The BLR consists of rapidly moving ($v \gtrsim 1000$ km~s$^{-1}$) gas. The strong variability of the lines in response to continuum variations shows that they arise almost entirely from photoionization and that other energy input is negligible.  Photoionization analyzes show that the gas is dense ($n_H \gtrsim 10^9$cm$^{-3}$).  I have recently reviewed the structure of the BLR elsewhere \citep{gaskell09} so I only give a brief sketch here.   From the large equivalent widths of the lines and the inferred shape of the ionizing UV continuum, we can calculate that the BLR gas must have a high covering factor of $\sim 40$\% \citep{gaskell+07}.  The absence of observed absorption from the BLR tells us that the BLR has a flattened distribution and that we must be viewing it from near the axis of symmetry \citep{antonucci+89,gaskell+07}.  BLR transfer functions (the responses to delta function continuum flares) also show that at least the low-ionization BLR has a flattened distribution with almost no material near the line of sight (\citealt{maoz+91,krolik+91,horne+91}). \citet{gaskell+07} point out that near the equatorial plane the covering fraction must be close to 100\%, so the BLR will be self-shielding. They show that this explains the strong observed radial ionization stratification.  \citet{gaskell+07} argue that the BLR will shield the torus as well so that the BLR and torus cover similar fractions of 4$\pi$ steradians.  The overall appearance of the torus and BLR can best be described as a bird's nest \citep{mannucci+92}.  A cartoon and visualizations can be found in \citet{gaskell+08} and \citet{gaskell09}.

\section{The direction of motion of the BLR}

Inflow, outflow, and virialized motions in either random orbits or more planar Keplerian orbits in a disk have all been considered as possibilities for BLR motions.  The presence of broad absorption lines (BALs) in some AGNs proves that {\em some} gas is outflowing.  Spatially-resolved outflow of some of the narrow-line region can also be seen.    However, BALs do {\em not} necessarily mean that the BLR is outflowing because BALs commonly extend to velocities {\em several times higher} than those observed for the BLR in the same objects (see, for example, \citealt{turnshek+88}).  It is thus not clear that there is necessarily any connection between BALs and BLRs.  I will argue here that BLR kinematics are consistent with the ``bird's'' nest geometry and that most of the BLR gas is {\em not} outflowing.

The question of the direction of motion can be settled by determining which sides of the black hole the blueshifted and redshifted wings of the line profiles arise from.  It has long been recognized that the most unambiguous way of doing this for emission-line gas is through velocity-resolved light echoes (e.g., \citealt{fabrika80,ulrich+84}) but early results were ambiguous.  Extensive ``reverberation mapping'' began with the introduction of the cross-correlation technique \citep{gaskell+sparke86,gaskell+peterson87} which has now been used to determine the effective radii of dozens of BLRs from line and continuum time series.  The technique can also be applied to the red and blue wings of a BLR line to determine the direction of motion \citep{gaskell88}.  If a line arises from a radiatively-driven outflow we can confidently predict that the peak of the cross-correlation function of the red-wing and blue-wing time series will show the blue wing leading the red by about twice the light-crossing time.  With only a couple of exceptions (discussed in section 5) this is never seen.

\citet{gaskell88} found that for NGC~4151 the wings of both the high-ionization C\,IV line and the low-ionization Mg\,II line varied almost simultaneously, but the {\em red} wings of both C\,IV and Mg\,II led the blue wings by $\sim 3$ and  $\sim 4.5$ days respectively.  Pure radial outflow was strongly excluded and virialized motion with significant inflow favored instead.  The C\,IV result was confirmed in a subsequent 1991 observing campaign when \citep{ulrich+horne96} found the red wing leading the blue by $1.95 \pm 0.03$ days. The profiles of Mg\,II and H$\beta$ generally agree in AGNs (\citealt{grandi+phillips79}) and their widths are well correlated from object to object (e.g., \citealt{gaskell+mariupolskaya02}), so we expect similar kinematics for the two lines.  \citet{maoz+91} found the red wings of both the broad H$\alpha$ and H$\beta$ lines to be leading their respective blue wings by 4 -- 5 days (see their Fig. 13) in good agreement with the \citet{gaskell88} Mg\,II result.

Similar behavior has been found in other objects.  \citet{koratkar+gaskell89} found the same combination of virialized motion plus inflow in Fairall~9.  For both C\,IV and Mg\,II pure outflow was again excluded at a high confidence level and the red wing of C\,IV led the blue wing by $75 \pm 45$ days (note that the BLR of Fairall~9 is more than 10 times bigger than that of NGC~4151).  Examination of {\it IUE} spectra of NGC~5548 taken over the 10-year period of 1978--1988 excluded pure radial outflow of the high-ionization BLR at $> 99$\% confidence \citep{koratkar+gaskell91a} and showed the red wing of C\,IV leading the blue wing by $2 \pm 5$ days.  Better sampled data from the 1988--1989 {\it International AGN Watch} ({\it IAW}) campaign showed the red wing leading by $5 \pm 3$ days \citep{crenshaw+blackwell90}.  Even better UV data from the 1993 {\it IAW} campaign \citep{korista+95} showed the red wing leading by $4.0 \pm 0.7$ days.  A detailed analysis of the same data set by \citep{done+krolik96} strongly supported the virialization-plus-inflow model and showed that the infall component of velocity of the high-ionization BLR at 5 light-days radius was fairly well constrained to be $1700 \pm 300$ km s$^{-1}$ (about a third of the line FWHM).  For H$\beta$ \citet{kollatschny+dietrich96} found the red wing leading the blue wing by $5 \pm 3$ days.  An analysis by \citet{welsh+07} of all {\it IAW} optical spectra from 1988-1993 showed the red wing of H$\beta$ leading the blue wing during this period by $5.0 \pm 1.3$ days.

\citet{koratkar+gaskell91b} also studied the red-wing/blue-wing lag for NGC~3783, NGC~4593, ESO~141-G55, Mrk~509, Mrk~926, and 3C~273.  The strong result from these early studies was that none of the AGNs showed the signature of a radiatively-accelerated outflow.  For most of the objects the sampling was too poor to be able to detect inflow signatures such as were seen in NGC~4151, Fairall~9 and NGC~5548, but for 3C 273 \citet{koratkar+gaskell91b} found the red wing of C\,IV leading the blue by $36 \pm 29$ days.  Even though in early studies the significance of the inflow signature was not often not high, the {\em combined} significance {\em was} high ($> 99$\%) and \citet{gaskell+snedden97} concluded that {\em all} objects were consistent with virialized motions plus infall.  This result has strengthened considerably with monitoring of more objects.  These include 3C~390.3 \citep{obrien+98,dietrich+98}, Mrk~6 \citep{sergeev+99}, Akn\,120 \citep{doroshenko+99,doroshenko+08}, Mrk~110 \citep{kollatschny03}, Mrk~40 \citep{bentz+08}, NGC~3227, and NGC~3516 \citep{denney+09}, and Mrk~1310, NGC~4748, and NGC~6814 \citep{bentz+09}.  For 3C~390.3 the red wings of Ly$\alpha$ and C\,IV led the corresponding blue wings by 19 and $17 \pm 11$ days respectively.  For Mrk~6 the red wing of H$\beta$ led by $14.5 \pm 3.5$ days.  For Akn\,120 \citet{doroshenko+99} give an 88\% confidence that the red wing of H$\beta$ leads the blue.  For Mrk~110 \citet{kollatschny03} the wings at $\pm 2400$ km~s$^{-1}$ show the red sides leading the blue by $\sim 13$ days for H$\beta$, $\sim 5$ days for He\,I $\lambda$5876, and $\sim 4$ days for He\,II $\lambda$4686.  For Mrk~40 the red side of H$\beta$ led by $4 \pm 1$ days, and for NGC~3516 it led by $\sim 6$ days.

As well as these many reports of the red wings of lines varying slightly before the blue wings, comparison of the wings and cores of lines shows that, as is expected for virialized motions, the high-velocity gas responds before the low-velocity gas \citep{clavel+91}.  Fig.\@ 1 shows the velocity--delay function from two years of monitoring of Akn\,120 by \citet{sergeev+99}.  One can see both that the red wing leads and that the wings vary faster than the core.  Almost identical velocity--delay functions can be found for Mrk~40 \citep{bentz+08} and NGC~3516 \citep{denney+09}.

\begin{figure}[t!]
\begin{center}
\epsfxsize = 70 mm
\epsfbox{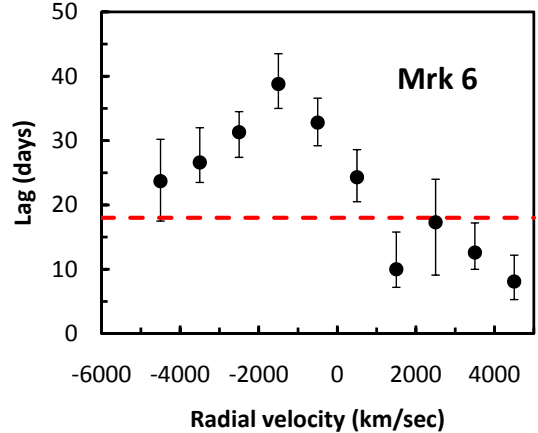}
\caption{The lag relative to the optical continuum of H$\beta$ emission in Mrk~6 as a function of radial velocity for 1993--1995.  Note (a) that the red wing responds before the blue wing, and (b) that the high-velocity gas at both positive and negative velocities responds faster than the slower velocity gas.  The dashed line shows the lag of the whole line.  Data from \citet{sergeev+99}.}
\label{CIV_NGC_4151}
\vspace{-0.5cm}
\end{center}
\end{figure}

\section{Accretion and the blueshifting of high-ionization lines}

\citet{gaskell82} discovered that the high-ionization C\,IV line was blueshifted with respect to the low-ionization BLR lines and proposed that this was a result of the high-ionization lines arising in a wind while the low-ionization lines arose from a disk hiding the receding side of the wind.  Strong blueshiftings of high-ionization lines have subsequently been widely interpreted as evidence of strong winds (e.g., in Narrow-Line Seyfert 1 galaxies (NLS1s)-- see \citealt{leighly+moore04}).  The big problem with the disk-wind explanation, however, is that {\em velocity-resolved reverberation mapping of C\,IV shows no evidence of outflow}.  \citet{gaskell+goosmann08} show that blueshifting from scattering off infalling material solves this problem, and that the observed infall velocities and geometry naturally explain the shifted profiles. It has long been recognized that if the BLR is inflowing, the mass influx is sufficient to account for AGN energy generation \citep{padovani+rafanelli88}.  In NLS1s the accretion is higher so the BLR mass influx must be higher.  This will produce a stronger blueshifting of the high-ionization lines \citep{gaskell+goosmann08}, as is observed.

\section{The limitations of reverberation mapping}
\vspace{-0.1cm}

Despite the substantial evidence for BLR infall in addition to virialized motion, a significant fraction of monitoring campaigns have {\em not} shown a clear infall signature.  Examples include NGC~4593 \citep{kollatschny+dietrich97}, NGC~3227 and NGC~5548 \citep{denney+09}, Mrk~1310, NGC~4748, and NGC~6814 \citep{bentz+09}.  The 1989 NGC~5548 monitoring reveals what is going on.  For the whole campaign \citet{kollatschny+dietrich96} found the same infall signature for the Balmer lines that \citet{crenshaw+blackwell90} had found for C\,IV, but {\em for the first outburst alone}, both H$\alpha$ and H$\beta$ showed a strong {\em outflow} signature!  Given this, it is not surprising that \citet{welsh+07} found substantially different inflow signatures for different observing seasons.  For example, in 1990 the blue/red lag was $0.0 \pm 1.3$ days while in 1992 it was $5.8 \pm 1.1$ days.

Since the two roughly comparable 1989 outbursts were separated by only 100 days (comparable to the light-crossing time) it is ludicrous to think that the direction of motion of the entire BLR changed!  Instead, the changes in kinematic signature are showing us the limits of reverberation mapping and caution us that when we see an outflow signature in an AGN in a single short observing campaign, (e.g., NGC~3227; \citealt{denney+09}), it is does {\em not} mean that the BLR is outflowing.  Conversely, a very strong {\em inflow} signature (e.g., in Mrk~40; \citealt{bentz+08}) does not necessarily mean strong inflow.

I believe the problems arise because real AGNs are messy objects!  Different parts of line profiles show different responses to observed continuum variability (e.g., \citealt{sergeev+01,shapovalova+04}).  For Mrk~6 \citet{sergeev+99} pointed out that some changes in the Balmer line profiles shapes cannot be caused by matter redistribution or light-travel time effects, but are probably caused by changes in the anisotropy of the ionizing continuum.  An anisotropic ionizing continuum could arise in several ways.  \citet{gaskell+klimek03} and \citet{gaskell06} argue that variable components of AGN continuum emission at high energies are likely to be intrinsically anisotropic.  Even if the intrinsic variability is not anisotropic, X-ray observations show that large variations in the absorbing column are common (\citealt{risaliti+02}; see \citealt{turner+miller09} for an extensive review) -- especially when viewing the system near edge on, as the BLR is doing.  These changes can be large and rapid.  NGC~1365, for example, changed from being Compton-thick to Compton-thin \citep{risaliti+05} in only a few weeks!  There will be a huge effect on the response of a region of BLR gas as optically-thick  material passes through its line of sight to an ionizing continuum source.  A third source of anisotropy is non-axisymmetric variability of the disk which has already been proposed as a cause of differing reverberation responses to differing outbursts \citep{gaskell08}.

I believe that it is these deviations from the standard reverberation-mapping assumption of an isotropic, centrally-located ionizing continuum source that are now limiting reverberation mapping campaigns rather than poor sampling or poor signal-to-noise ratios .  In particular {\em there is a significant risk of getting erroneous reverberation mapping results from observing campaigns of short duration where the results are dominated by single events.}  It is more important to have {\em longer} monitoring campaigns rather than denser sampling.

\begin{acknowledgments}
I am grateful to Ren\'e Goosmann and Ski Antonucci for discussions, and to the NSF for support through grant AST 08-03883.
\end{acknowledgments}

\end{document}